\documentclass[twocolumn,showpacs,preprintnumbers,amsmath,amssymb]{revtex4}

\usepackage{psfig}
\usepackage{graphicx}
\usepackage{dcolumn}
\usepackage{bm}

\def\be{\begin{equation}}
\def\ee{\end{equation}}
\def\bea{\begin{eqnarray}}
\def\eea{\end{eqnarray}}

\begin{document}

\title{Optical control of spin coherence in singly charged (In,Ga)As/GaAs quantum dots}

\author{A. Greilich$^1$, R. Oulton$^1$, E.~A. Zhukov$^{1,*}$, I.~A. Yugova$^{1,\dag}$ , D.~R. Yakovlev$^{1,3}$, M. Bayer$^1$,
A. Shabaev$^{2,\ddag}$, Al.~L. Efros$^2$, I.~A. Merkulov$^3$, V.
Stavarache$^4$, D. Reuter$^4$, and A. Wieck$^4$}

\affiliation{$^1$ Experimentelle Physik II, Universit\"at
Dortmund, D-44221 Dortmund, Germany} \affiliation{$^2$ Naval
Research Laboratory, Washington, DC 20375, USA} \affiliation{$^3$
A.~F. Ioffe Physico-Technical Institute, RAS, St. Petersburg,
194021 Russia} \affiliation{$^4$ Angewandte Festk\"orperphysik,
Ruhr-Universit\"at Bochum, D-44780 Bochum, Germany}

\date{\today}

\begin{abstract}
Electron spin coherence has been generated optically in $n$-type
modulation doped (In,Ga)As/GaAs quantum dots (QDs) which contain
on average a single electron per dot. The coherence arises from
resonant excitation of the QDs by circularly-polarized laser
pulses, creating a coherent superposition of an electron and a
trion state. Time dependent Faraday rotation is used to probe the
spin precession of the optically oriented electrons about a
transverse magnetic field. Spin coherence generation can be
controlled by pulse intensity, being most efficient for
$(2n+1)\pi$-pulses.
\end{abstract}
 \maketitle

An electron spin in a single QD represents a qubit candidate that
is very attractive for solid state quantum information processing
\cite{LossPRA98,ImamogluPRL99,Spintr}, as has been suggested by
long electron spin coherence times, $T_2$, measured in bulk
semiconductors. \cite{Kikkawa} These long times are required for
performing a sufficient number of quantum manipulations during
which coherence needs to be retained. Recent QD studies have
demonstrated long electron spin relaxation lifetimes, $T_1$, in
the millisecond-range at cryogenic temperatures.
\cite{Elzermannature04} This has raised hopes that $T_2$, which
may last as long as $2T_1$, \cite{Loss2004} could be similarly
long, with encouraging indications to that effect found lately.
\cite{science} For fast spin manipulation, rotations by Raman
processes are envisaged whose cross sections can be enhanced by
resonant excitation of a charged exciton. \cite{ChenPRB03} In a
first step, however, electron spin coherence must be established,
which recently was addressed in charged GaAs/AlGaAs interface QDs.
\cite{DuttPRL05} However, only rather low excitation powers were
used in those experiments, so that coherent control of electron
spin polarization in form of Rabi oscillations did not occur.

In this Letter we demonstrate by pump-probe Faraday rotation (FR)
that electron spin coherence can be generated by circularly
polarized optical excitation of singly charged QDs. Resonant
excitation creates an intermediate superposition of a singlet
trion and an electron, which after trion radiative decay is
converted into a long lived electron spin coherence. The coherence
is controlled by the pump pulse area, $\propto \int E(t)dt$, where
$E(t)$ is the electric field amplitude. It reaches maximum for
$(2n+1) \pi$-pulses \cite{Scully}, in good accord with our
theoretical model. Theory also shows that $2n \pi$-pulses can be
used for refocussing of the precessing spins.

\begin{figure}[tbp]
\centering \centerline{\psfig{figure=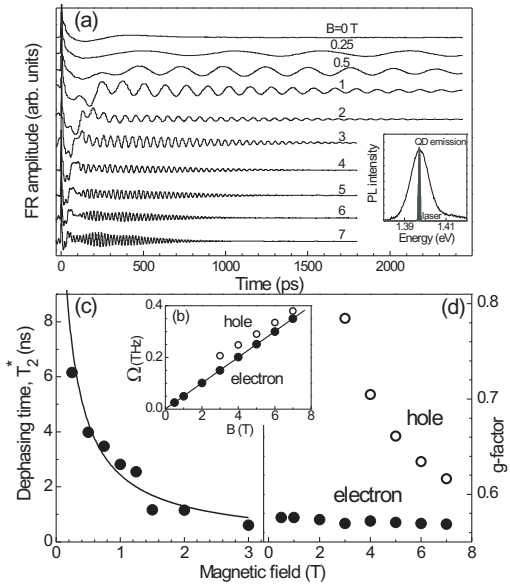,width=8cm}}
\caption{(a): FR traces of $n$-doped (In,Ga)As/GaAs QDs vs delay
between pump and probe at different $B$. Pump power was $\sim$ 10
mW. Inset: Photoluminescence spectrum of these QDs compared to
laser spectrum in FR studies.  (b) and (d): Field dependencies of
electron/hole precession frequencies and electron and hole
$g$-factors. (c): Spin dephasing time $T_2^{\star}$ vs $B$
(symbols).  Line is a $1/B$-fit to data.} \label{fig1}
\end{figure}

The experiments were performed on self-assembled (In,Ga)As/GaAs
QDs. To obtain strong enough light-matter interaction, the sample
contained 20 layers of QDs separated by 60 nm wide barriers. It
was fabricated by molecular beam epitaxy on a (001)-oriented GaAs
substrate. The layer dot density is about 10$^{10}$ cm$^{-2}$. For
an average occupation by a single electron per dot, the structures
were $n$-modulation doped 20 nm below each layer with a Si-dopant
density roughly equal to the dot density. The sample was thermally
annealed so that its emission occurs around 1.396 eV, as seen from
the luminescence spectrum in the inset of Fig. \ref{fig1}(a). The
full width at half maximum of the emission line is 10 meV,
demonstrating good ensemble homogeneity. Further optical
properties of these dots and undoped reference sample can be found
in Ref. \cite{Gre06}.

The sample was immersed in liquid helium at temperature $T$ =
2\,K. The magnetic field $B \leq$ 7 T was aligned perpendicular to
the structure growth direction $z$. In the FR pump-probe studies
\cite{Spintr} a Ti-sapphire laser emitting pulses with a duration
of $\sim$1 ps (full width at half maximum of $\sim$2 meV) at 75.6
MHz repetition rate was used, hitting the sample along $z$. The
laser was tuned to the QD ground state transition energy (see
inset of Fig. \ref{fig1}(a)). The circular polarization of the
pump beam was modulated at a frequency of 50 kHz, to avoid nuclear
polarization effects. For detecting the rotation angle of the
linearly polarized probe pulses, a homodyne technique based on
phase-sensitive balanced detection was used.

Figure \ref{fig1}(a) shows the FR signal of the QDs vs delay
between pump and probe for different magnetic fields. Pronounced
electron spin quantum beats are observed with some additional
modulation at high $B$. With increasing delay time the beats
become damped. The oscillations at low $B$ last much longer (for
example, about 4 ns at 0.5 T) than the radiative trion lifetime of
$\tau_r= 400$ ps, as measured by time-resolved photoluminescence
and therefore are due to long-lived residual electrons. Three
features are to be noted:

1) The oscillation frequency increases with magnetic field as
expected from the spin-splitting of electron states: $\hbar
\Omega_e=g_{e}\mu_B B$, where $g_{e}$ is the electron $g$-factor
and $\mu_B$ is the Bohr magneton. We have analyzed the FR dynamics
in Fig. 1(a) by an oscillatory function with exponentially damped
amplitude, $\propto \exp \left( - t / T_2^{\star} \right) \cos
\left( \Omega_e t \right)$. The resulting $B$-field dependence of
the electron precession frequency is shown in Fig. 1(b). From a
$B$-linear fit we obtain $\mid g_{e} \mid = 0.57$.

2) The spin beats become increasingly damped with increasing
magnetic field, corresponding to a reduction of the ensemble spin
dephasing time $T_2^{\star}$, plotted in Fig. 1(c). The damping of
the spin precession arises from variations of the electron
$g$-factor within the QD ensemble, causing an enhanced spread of
$\Omega_e$ with increasing $B$, whose impact on the dephasing time
can be described by $\left[ T_2^{\star} (B) \right]^{-1} = \left[
T_2^{\star} \left(0\right) \right]^{-1}  + \Delta g_{e} \mu_B B /
\sqrt{2} \hbar$. The solid line in Fig. 1(c) shows a 1/$B$ fit to
the experimental data for $T_2^{\star}$, from which a g-factor
variation of $\Delta g_{e} = 0.004$ can be extracted. From the
data one can also conclude that $T_2^{\star}\left(0\right)$ is
longer than 6 ns. The zero-field dephasing is mainly caused by
electron spin precession about the frozen magnetic field of the
dot nuclei in a QD. \cite{Merkulov} The net orientation of nuclei
varies from dot to dot, and it is these variations that lead to
ensemble spin dephasing.

\begin{figure}[tbp]
\centering \centerline{\psfig{figure=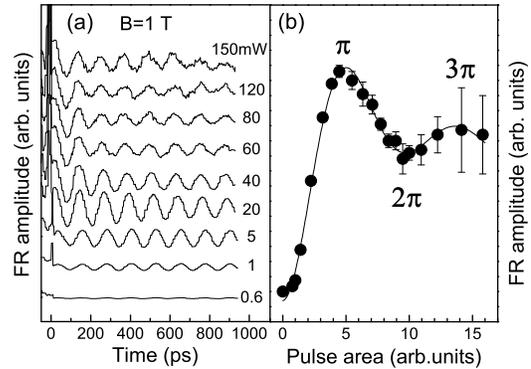,width=7cm}}
\caption{(a): Closeup of FR rotation signal at $B$ = 1\,T for
different pump powers.  (b): FR amplitude vs laser pulse area.
Line is guide to the eye.} \label{fig2}
\end{figure}

3) The additional modulation of the FR traces at high fields is
observable only during the trion lifetime and therefore can be
naturally assigned to the photoexcited holes with a spin-splitting
$\hbar \Omega_h=g_{h}\mu_B B$, where $g_{h}$ is the hole
$g$-factor. The analysis gives $\mid g_h \mid$=0.66 at $B$=5 T.
The decay time of this mode is 170 ps which is significantly
shorter than the 400 ps decay of the electron spin precession for
this field strength, pointing either at strong variations of $g_h$
or at fast hole spin  relaxation. The $B$-dependence of $\Omega_h$
is shown in Fig. \ref{fig1} (b), and the hole g-factor vs $B$ is
given in Fig. \ref{fig1}(d). $g_h$ varies with $B$ over the
studied field range, while $g_e$ is constant. In the range from 3
to 7 T, where the beat modulation is detectable, $g_h$ decreases
from 0.78 to 0.62. This dependence suggests that mixing of
light-hole and heavy-hole states is significant for the studied
QDs. \cite{Henneberger}

Figure \ref{fig2}(a) shows FR signals at $B$ = 1\,T for different
pump powers. The corresponding FR amplitude is plotted in Fig.
\ref{fig2}(b) as function of the pulse area, which is defined as
$\Theta = 2\int \left[\mbox{\boldmath{$d E$}({\it t})}\right]dt /
\hbar$ in dimensionless units, where $\mbox{\boldmath$d$}$ is the
dipole transition matrix element. For pulses of constant duration,
but varying power, as used here, the pulse area is proportional to
the square root of excitation power, and it is given in arbitrary
units in Fig. \ref{fig2}(b). The amplitude shows a non-monotonic
behavior with increasing pulse area. It rises first to reach a
maximum, then drops to about 60\%. Thereafter it shows another
strongly damped oscillation. This behavior is very similar to the one known from Rabi-oscillations of the 
Bloch vector for varying excitation power.\cite{Rabi} The FR
amplitude becomes maximum when applying a $\pi$-pulse as pump, 
for which the $z$-component of the Bloch vector is fully inverted.
It becomes minimum for a $2 \pi$-pulse, for which the Bloch vector
is turned by 360$^{\circ}$, and so on. The damping of the Rabi
oscillations most likely is due to ensemble inhomogeneities of QD
properties such as the transition dipole moment. \cite{Borri}
These observations are important input for identifying the origin
of spin coherence.

\begin{figure}
\centering \centerline{\psfig{figure=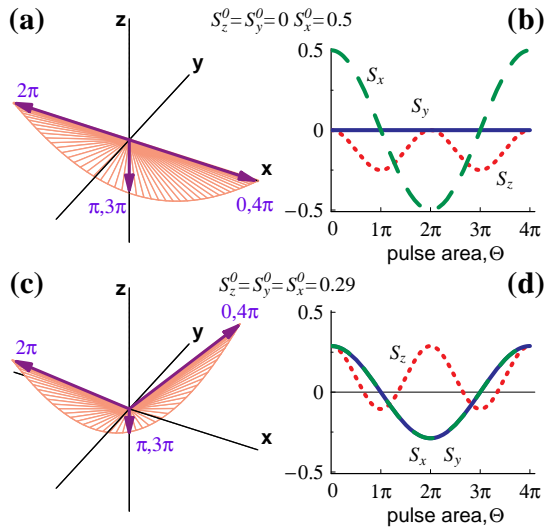,width=7.5cm}}
\caption{(a) and (c): Reorientation of electron spin polarization
by application of a resonant optical pulse of varying area as
denoted. Calculations have been done for two different initial
values of spin polarizations, $S_x^0$ and $S_z^0$. (b) and (d):
Electron spin polarization components vs pulse area $\Theta$.}
 \label{fig3}
\end{figure}

This origin will be discussed in the following: The magnetic field
$\mbox{\boldmath $B$} \parallel \mbox{\boldmath $e$}_x$  leads to
a splitting of the electron spin into eigenstates with spin
parallel ($+x$) or antiparallel ($-x$) to the field. Disregarding
the hole for simplicity, for {\sl neutral} QDs the optical pulses
create electrons with spin states $|\uparrow\rangle$ or
$|\downarrow\rangle$ along the $z$-direction of light propagation,
$S_{z} = \pm 1/2$, which can be expressed as coherent
superpositions of the two spin eigenstates $\mid \pm x \rangle$.
Therefore spin quantum beats occur, which in a classical picture
can be treated as electron spin precession about the magnetic
field. In any case, the spin precession cannot last longer than
the exciton lifetime.

In {\sl charged} QDs, resonant optical excitation leads to
formation of trions $|\uparrow\downarrow\Uparrow\rangle$ or
$|\uparrow\downarrow\Downarrow\rangle$, consisting of two
electrons which form a spin singlet and a hole in one of the
states $|\Uparrow\rangle$ or $|\Downarrow\rangle$ with spin
projection $J_{h,z} = \pm 3/2$ for $\sigma^+$ or $\sigma^-$
polarized light. \cite{Tischler} The resident electron may have
arbitrary spin orientation
$|\psi\rangle=\alpha|\uparrow\rangle+\beta|\downarrow\rangle$,
where $|\alpha|^2+|\beta|^2=1$, if the trion state is not
populated, and $|\alpha|^2+|\beta|^2<1$, otherwise. The electron
spin polarization is characterized by the spin vector
$\mbox{\boldmath $S$} = (S_x, S_y, S_z)$ which can be introduced
as: $S_x={\rm Re}(\alpha\beta^*)$, $S_y=-{\rm Im}(\alpha\beta^*)$, and
$S_z=(1/2)(|\alpha|^2-|\beta|^2)$. Similarly, one can introduce
the trion spin vector, $\mbox{\boldmath $J$} = (J_x, J_y, J_z)$,
that represents the polarization of the trion,
$|\bar{\psi}\rangle=\bar{\alpha}|\uparrow\downarrow\Uparrow\rangle+
\bar{\beta}|\uparrow\downarrow\Downarrow\rangle$.

A {\em short} pulse of circularly polarized light is a remarkable
tool for controlling coherently an electron spin in a transverse
magnetic field. If the pulse length, $\Delta t$, is much shorter
than the radiative decay time and electron and hole spin
relaxation times, it  mixes the electron and trion spin states
into a superposition state, which is not affected by the
corresponding decoherence processes. For controlled resonant
pumping as in experiment, the coherent superposition is uniquely
determined by the pulse area $\Theta$. By variation of this area
not only the electron and trion state populations can be changed
periodically with period $\Theta=2\pi$, but even more important
also the orientation of the electron and trion spins
$\mbox{\boldmath $S$}$ and $\mbox{\boldmath $J$}$ can be
controlled.

The three dimensional spin vectors $\mbox{\boldmath $S$}$ and
$\mbox{\boldmath $J$}$ represent six of the sixteen components of
the four level density matrix and their dynamics can be described
by density matrix equations of motion. \cite{SEtoBEpub} The
evolution of the electron spin vector as function of $\Theta$ is
shown in Fig. \ref{fig3} for two initial spin directions: one is
parallel to the magnetic field and the other exemplifies an
arbitrary direction. A short $\sigma^+$ polarized pulse resonantly
excites the initial electron spin state, $|\psi_0\rangle =
\alpha_0|\uparrow\rangle + \beta_0|\downarrow\rangle$, into the
coherent superposition of the electron and trion states
$|\psi_{\rm ET}\rangle=\alpha^0\cos(\Theta/2)|\uparrow\rangle  +
\beta^0|\downarrow\rangle -
i\alpha^0\sin(\Theta/2)|\uparrow\downarrow\Uparrow\rangle$. The
light induced deviation of the $S_z$ component, $\left|S_z -
S_z^0\right| = |\alpha_0|^2\sin^2(\Theta/2)$ changes with the
$|\uparrow\rangle$ state population, and independently of the
initial conditions it reaches a maximum for $(2n+1)\pi$-pulses
with $\Theta=(2n+1) \pi$, for which the $S_x$ and $S_y$ components
vanish. In particular, $S_z([2n+1]\pi)=-0.25$ for $S_z^0=0$, in
agreement with Ref. \cite{SEM}. It is interesting to note that,
unlike the $S_z$ component, the electron spin, swings between the
initial direction $(S_x^0, S_y^0, S_z^0)$ and the direction
$(-S_x^0, -S_y^0, S_z^0)$ with a period $4\pi$. This is because
the $S_{x,y}\sim \cos(\Theta/2)$ components describe the coherence
of the electron spin state and both components change with the
phase of the spin wave function, $|\psi_{\rm ET}\rangle$.

The control of spin dynamics by an optical pulse allows for a fast
spin alignment. In a QD ensemble, a small area pulse, $\Theta \ll
1$, induces a coherent spin polarization proportional to $\Theta$,
as observed in shallow GaAs/AlGaAs interface QDs Ref.
\cite{DuttPRL05}. With increasing $\Theta$, the total spin
polarization oscillates as does the $S_z$ component of each
individual spin in the ensemble all of which oscillate with the
same period, $2\pi$. The long trion lifetimes in our QDs could
enable realization of regime in which pulse of rather low power,
but long duration can be used to reach these large pulse area
without decoherence due to radiative decay. This explains the FR
amplitude oscillations with pulse area in Fig. \ref{fig2}.
Further, the $S_x$ and $S_y$ components change sign with period
$2\pi$. This implies that $2n\pi$-pulses can be used for
refocusing of precessing spins, very similar to spin-echo
techniques. \cite{p-puls}


Let us turn now to the spin dynamics after its initialization by
the short pulse: The trion component of the electron-trion
coherent superposition decays spontaneously leading to the
observed long-lived spin precession of the resident electron.
After the pulse, the off-diagonal component of the density matrix,
describing electron-trion coherence becomes decoupled from the
electron and trion spin vectors, $\mbox{\boldmath $S$}$ and
$\mbox{\boldmath $J$}$, which are governed independently by two
coarse-grained vector equations: \bea
  {d \mbox{\boldmath $J$}\over dt}&=&[\mbox{\boldmath
  $\Omega$}_h\times \mbox{\boldmath $J$}]-{\mbox{\boldmath $J$}\over \tau_s^h}-
  {\mbox{\boldmath $J$}\over \tau_r},\nonumber\\
  {d \mbox{\boldmath $S$}\over dt}&=&[(\mbox{\boldmath
  $\Omega$}_e+\mbox{\boldmath
  $\Omega$}_N)\times \mbox{\boldmath $S$}]+ {\left(\hat{\mbox{\boldmath $J$}} \hat{\mbox{\boldmath
  $z$}}\right) \hat{\mbox{\boldmath $z$}}\over\tau_r},
\eea where $\mbox{\boldmath $\Omega$}_{e,h} \parallel
\mbox{\boldmath $e$}_x$ and $\mbox{\boldmath
$\Omega$}_{N}=g_e\mu_B\mbox{\boldmath $B$}_{N}/\hbar$ is the
precession frequency of the electron in an effective nuclear
magnetic field, $\mbox{\boldmath $B$}_{N}$. We do not include in
the second equation the electron spin relaxation time,
$\tau_s^{e}$, in an explicit form. At low temperatures,
$\tau_s^{e}$ is on the order of $\mu$s and it is mainly determined
by fluctuations of the nuclear field $\mbox{\boldmath
$\Omega$}_{N}$ in a single QD \cite{Merkulov,KhaetskiiPRL02}.
This time
scale is irrelevant to our present problem. 
The spin
relaxation of the hole in the trion, $\tau_s^h$, is caused by
phonon assisted processes and  at low temperatures it may be as
long as $\tau_s^{e}$ in QDs. \cite{LossPRL}

Solving Eq.(1) we obtain the time evolution of the spin vectors
$\mbox{\boldmath $S$}$ and $\mbox{\boldmath $J$}$. After trion
recombination ($t \gg \tau_r $), the amplitude of the long-lived
electron spin polarization excited by a $(2n+1)\pi$-pulse is given
by \bea
  S_z\left(t \right) &=& {\rm Re}
    \left\{
      \left(
        S_z(0) +
        {0.5J_z(0)/\tau_r\over \gamma_T + i\left(\omega + \Omega_h\right)}\right.\right.\nonumber\\
        &+&\left.\left.
        {0.5J_z(0)/\tau_r\over \gamma_T + i\left(\omega - \Omega_h\right)}
      \right)\exp(i\omega t)
  \right\},
\eea where  $S_z(0)$ and $J_z(0)$ are the electron and  trion spin
polarizations created by the pulse. $\omega=\Omega_e+
\Omega_{N,x}$. $\gamma_T=1/\tau_r+1/\tau_s^h$ is the total trion
decoherence rate. On average, the induced spin polarization
$S_z(t)$ is nullified by trion relaxation, as $S_z(0)= -J_z(0)$,
if the radiative relaxation is fast $\tau_r \ll
\tau_s^h,\,\Omega_{e,h}^{-1}$. In contrast, if the spin precession
is fast, $\Omega_{e,h} \gg \tau_r^{-1}$, the electron spin
polarization is maintained after trion decay \cite{SEM,Economou},
as observed in our case.

\begin{figure}
  \centering
\centerline{\psfig{figure=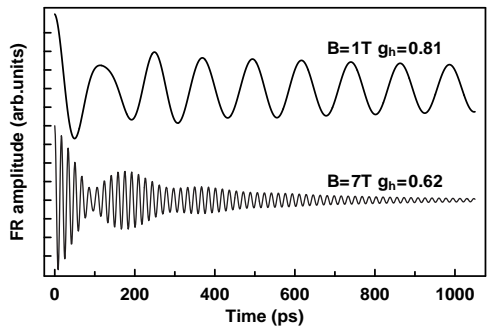,width=7cm}}
\caption{Calculated time dependence of pump-probe FR signal of
$n$-doped QDs excited by a $\sigma^+$ polarized pulse.
$\tau_r$=400 ps, $\tau_s^h$=170 ps, $\mid g_{e} \mid = 0.57$, and
$\Delta g_{e} = 0.004$ .} \label{fig4}
\end{figure}

For an ensemble of QDs, the electron spin polarization is obtained
by averaging Eq.~(1) over the distribution of $g$-factors and
nuclear configurations. At low magnetic fields, the random
magnetic field of nuclei becomes more important for the electron
spin dephasing than $g$-factor dispersion, leading to dephasing
during several nanoseconds. \cite{Merkulov} The rotation of the
linear probe polarization is due to the difference in scattering
of the $\sigma^+$ and $\sigma^-$ polarized components by one of
the allowed transitions
$|\uparrow\rangle\rightarrow|\uparrow\downarrow\Uparrow\rangle$
and
$|\downarrow\rangle\rightarrow|\uparrow\downarrow\Downarrow\rangle$.
The scattering efficiency is proportional to the population
difference of the states involved in the transition $\Delta n_+ =
n_\uparrow - n_\Uparrow$ or $\Delta n_- = n_\downarrow -
n_\Downarrow$, and the FR angle is $\phi(t)\sim(\Delta n_+ -
\Delta n_-)/2 = S_z (t) - J_z (t)$.

Figure \ref{fig4} shows the FR signal after a $\sigma^+$-polarized
excitation pulse, calculated with input parameters from
experiment. At $B=7$\,T, the FR shows modulated beats resulting
from interference of the electron and trion precessions during the
400 ps trion lifetime. At $B=1$\,T the beats are less pronounced
because of the larger difference between electron and hole
$g$-factors. These results are in good agreement with experiment.

In conclusion, we have shown experimentally and theoretically that
pulses of circularly polarized light allow for a coherent phase
control of an electron spin in a QD. The coherent control results
in FR amplitude oscillations with varying laser pulse area.

{\bf Acknowledgments} This work was supported by the DARPA program
QuIST, ONR, CRDF, DFG,
and the BMBF program 'nanoquit'. R. O. thanks the Alexander von
Humboldt foundation.

\end{document}